\documentclass[epj]{svjour}
\usepackage{graphics}
\begin{document}
\title{Flavour asymmetry of anti-quarks in nuclei}
\author{Koichi Saito % etc
\thanks{e-mail: ksaito@nucl.phys.tohoku.ac.jp}%
}                     % Do not remove
\institute{Tohoku College of Pharmacy, Sendai 981-8558, Japan}
\date{Received: date / Revised version: date}
\abstract{
A novel nuclear effect on the flavour asymmetry of anti-quarks in the 
nucleon bound in a nucleus is discussed in terms of the meson cloud 
model and the Pauli exclusion principle. It is expected that the flavour 
asymmetry is enhanced in a nucleus. 
\PACS{
      {13.60.Hb}{Nuclear structure functions (flavour asymmetry)}   \and
      {21.90.+f}{Mesons in nuclei}
     } % end of PACS codes
} %end of abstract
\maketitle
It is a rather surprising that the distribution of $\bar{u}$ is different 
from that of $\bar{d}$ in the free proton ($p$) because 
those distributions are 
expected to be almost flavour symmetric within perturbative quantum 
chromodynamics (pQCD). The measurement of the flavour nonsinglet 
structure function, $F_2^p(x,Q^2) - F_2^n(x,Q^2)$, where $F_2^{p(n)}(x,Q^2)$ 
is the proton (neutron ($n$)) structure function, performed 
by the New Muon Collaboration (NMC)~\cite{nmc} led to a large ($\sim 30 \%$) 
violation of Gottfried sum rule (GSR), which implies that the distribution 
of $\bar{d}$ overcomes that of $\bar{u}$ in $p$. Later, this was confirmed 
by the Drell-Yan and semi-inclusive measurements of NA51~\cite{na51}, 
E866 (NuSea)~\cite{e866} and HERMES~\cite{hermes} collaborations. 

So far, various models have been proposed to understand the flavour asymmetry 
in $p$. Among those the meson cloud model~\cite{meson,rev} is successful in 
explaining the observed results. The physical proton contains many virtual 
meson-baryon components, and the valence anti-quark in the meson can 
contribute to the anti-quark distributions in the proton sea: 
\begin{eqnarray}
|p\rangle_{phys} &=& Z^{1/2}|p\rangle_{bare} 
+ f_{\pi^0p} |\pi^0\rangle |p\rangle_{bare} 
+ f_{\pi^+n} |\pi^+\rangle |n\rangle_{bare} \nonumber \\
&+& f_{\pi^-\Delta^{++}} |\pi^-\rangle |\Delta^{++}\rangle_{bare} 
+ \cdots , 
\label{physp}
\end{eqnarray}
where $Z^{1/2}$ is the renormalization constant for the wave function, 
$f_{MB}$ stands for 
the amplitude of Fock component containing a meson $M (= \pi, \rho, 
\cdots)$ and a baryon $B (= p, n, \Delta, \cdots)$, and $|B\rangle_{bare}$ is 
the bare baryon state. Note that the pion 
cloud provides the largest contribution because of its small mass.  Since the 
probability of the $\pi^+$-$n$ Fock component is larger than that of the 
$\pi^-$-$\Delta^{++}$ state in $p$, a surplus of $\bar{d}$ is 
naturally explained in the meson cloud model. 

An alternative explanation for an excess of $\bar{d}$ over $\bar{u}$ in 
$p$ involved the Pauli exclusion principle at quark level~\cite{rev}, 
given that there are two valence $u$ quarks in the proton and one valence 
$d$. In a model such as the bag model, where the quarks are confined by a 
scalar potential, the vacuum inside a hadron is different from the 
vacuum outside. This manifests itself as a distortion in the Dirac sea, 
which is full outside, whereas there will be empty states inside the hadron. 
To an external probe this change in vacuum structure appears as an intrinsic, 
non-perturbative sea of $q{\bar q}$ pairs even in the bare proton state 
(see Eq.(\ref{physp}) and Ref.~\cite{chiral}). 
Hence, because of Pauli blocking the presence of a valence quark in the 
hadron ground state lowers the 
probability of a negative energy state being empty, which is the same as 
lowering the probability of finding a positive energy anti-quark. 
If the number of valence $u$ quarks is equal 
to that of $d$ quarks in the hadron ground state, no flavour asymmetry 
appears. However, as the proton 
consists of two valence $u$ and one valence $d$ quarks, the asymmetry is 
realized and it is expected to be ${\bar d}/{\bar u} \sim 5/4$ from a naive 
counting estimate in the free proton~\cite{rev}. 
It may well be that the experimentally observed excess involves 
contributions from both of those effects. 

It is a very interesting problem to see how the flavour asymmetry changes in 
a nucleus.  Such a study would give us more information on the 
non-perturbative structure of the nucleon. Recently we have 
proposed~\cite{sai1,sai2,sai3} that the flavour asymmetry in a nuclear 
medium can be investigated by measuring 
the nonsinglet difference between structure 
functions of a pair of mirror nuclei $(A, A')$, where $A = Z + N (Z > N)$ 
(proton rich) and $A' = Z' + N' (N' > Z')$ (neutron rich) with $Z$ the 
proton number and $N$ the neutron number. 

How is the meson field modified in a nucleus ? To consider it we concentrate 
here on only Fock states involving the bare nucleon and pion. (See again 
Eq.(\ref{physp}). The admixture of $\pi$-$N$ component in the physical 
proton is about 
20\%, while those for $\rho$-$N$ and $\pi$-$\Delta$ are respectively about 
10\% and 5\%~\cite{rev}.)  First, 
we suppose that there is {\em no} charge symmetry breaking (CSB) 
in a nucleus. Then, in the proton rich nucleus $A$ the virtual 
emission of $\pi^- ({\bar u}d)$ from a neutron ($n \to \pi^- + p$) is 
{\em more suppressed} than the $\pi^+ (u{\bar d})$ from $p$ 
($p \to \pi^+ + n$) 
because of the Pauli blocking effect on the proton in the final state.  
On the contrary, in $A'$ (neutron rich)  the $\pi^+$ 
emission is more suppressed than the $\pi^-$ emission.  Note that from the 
point of view of Pauli blocking 
the $\pi^0$ field is not changed a great deal in
a nulceus. 

This effect is quite similar to the Pauli exclusion effect at 
quark level for the free proton. Usually a nucleus can be regarded as a 
sum of a closed-shell, core nucleus and some valence nucleons. 
If we treat the core nucleus as a new vacuum, the difference 
between the numbers of valence $p$ and $n$ decides how the individual pion 
field in isospin space is modified in the nucleus. 

Taking the flavour nonsinglet combination of the structure functions, i.e., 
$F_2^A(x) - F_2^{A'}(x)$, one can study the flavour asymmetry in 
the {\em bound} proton and neutron.  
Because of the Pauli exclusion principle at hadron level, the 
flavour asymmetry in the bound proton is expected 
to be more enhanced than that for the free proton: 
$({\bar d} - {\bar u})_{bound} > ({\bar d} - {\bar u})_{free}$. This is
a novel nuclear effect on the flavour asymmetry of the anti-quarks, in
addition to the asymmetry observed in free space.
Note that if the number of valence $p$ is identical to that of $n$ the 
present asymmetry vanishes. 

It is not allowed to ignore the effect of CSB (mainly Coulomb effct) 
in reality. Owing to the Coulomb repulsion the energy levels for the protons 
are slightly shallowed, which helps to enhance the virtual emision 
of $\pi^+$ from a proton in $A$. In $A'$
this effect acts on the $\pi^-$ emission (from a neutron) oppositely. 
However, for light (stable) nuclei the Coulomb effect on the 
pion emission may be small compared with the Pauli blocking effect 
discussed above. 

There is an anticipated possibility of finding exotic configurations, 
such as 6-quark states in a nucleus. As a consequence, the effect of the 
Pauli exclusion principle at quark level is different from that for the 
free proton. In $A$ (proton rich) the possibility of finding a 6-quark 
state created from two protons~\cite{sai4} would be larger than that of a 
two-neutron 6-quark state. In such a case, 4 valence $u$ 
and 2 valence $d$ quarks are supposed to be put into one 
confining potential. 
The naive counting estimate then gives ${\bar d}/{\bar u} 
\sim 2$ in the two-proton 6-quark state.  In $A'$ the opposite situation 
would occur. Hence, we again expect that the excess of ${\bar d}$ over 
${\bar u}$ is enhanced in the proton bound in $A$. 

Experimentally there is {\em only} one pair of {\em stable} mirror nuclei, 
that is, 
$^3$He-$^3$H. In our recent work~\cite{sai3} we have discussed this $A=3$ 
system and reported that the change of the pion fields leads to an
enhancement of the asymmetry, which predicts a reduction of the (nuclear) 
GSR by about $10\%$ compared with the GSR value in free space. 
The pair of $^7$Li-$^7$Be is the next 
candidate, but unfortunately $^7$Be is unstable although its half-life 
time is long (53 days). For other candidates one can find several 
pairs of mirror nuclei.  But, because of short half-life 
times, it is nearly impossible for the present to measure those 
structure functions with fixed-target experiments. 

If deep-inelastic scattering with high energy electrons (or muons) off 
{\em unstable} nuclei is realized by using a collider machine~\cite{muses} 
in the future, available mirror pair would be widely extended~\cite{tani} 
and we could measure structure functions of unstable nuclei 
systematically~\cite{sai5}. 

In particular, a nucleus, which is in the 
vicinity of the neutron or proton drip line, is quite interesting -- such 
a nucleus is sometimes called a halo nucleus. For a stable 
nucleus ($Z \approx N$) the potentials for protons and neutrons are almost 
the same except that the protons see a shallower potential due to 
the Coulomb effect.  As a number of neutrons increases 
and hence the nucleus is closer to the neutron drip line, the potential 
felt by protons becomes deeper because of an attractive $p$-$n$ 
interaction. This phenomenon leads to an enhancement of the $\pi^-$ emission 
from a neutron in such a (neutron rich) halo nucleus ($A'$). On the contrary, 
in a (proton rich) halo nucleus ($A$) the $\pi^+$ emission from $p$ is 
enhanced due to the Coulomb effect (as discussed above). 
Hence, the flavour symmetry 
of anti-quarks in the proton bound in an 
unstable, proton rich nucleus would be more broken compared with 
the case of stable one.  At present we can find two candidates for 
mirror nuclei with halos: ($^8$B-$^8$Li) and 
($^{17}$Ne-$^{17}$N)~\cite{tani}. 

It is very interesting to study the flavour asymmetry in the spin structure 
function. Recently several people have calculated the polarized anti-quark 
distributions in the meson cloud model~\cite{pol1,pol2}. It is clear that the 
$\rho$ meson plays an important role in the spin structure function, but 
the situation seems to be still confusing about how it 
contributes to the spin asymmetry~\cite{pol1,pol2}. 
In the meson cloud model, as in the case of the pion field, 
the $\rho^+ (\rho^-)$ emission from a proton (neutron) in $A (A')$ would be 
more enhanced than that in free space. 
Therefore, the flavour asymmetry in the spin asymmetry, $\Delta{\bar d} - 
\Delta{\bar u}$, is again expected to be enhanced in the proton bound in $A$. 
However, Cao and Signal~\cite{pol2} have reported that in the free proton 
the Pauli blocking effect at quark level has an opposite sign to the 
$\rho$ meson contribution and that its magnitude is much larger than 
the meson cloud effect.  Thus, in the case of spin-dependent nuclear 
structure functions it would be difficult to measure an enhancement of 
the flavour asymmetry caused by the change of the $\rho$ fields in nuclei.  
Nevertheless, it is very intriguing to study the Bjorken 
sum rule for a pair of mirror nuclei (see, for example, Ref.~\cite{spin}). 

So far, we have restricted ourselves to only 
Fock states containing the nucleon and pion in the meson cloud model. We 
have then found that the Pauli blocking effect plays an important role in the 
process of a virtual pion emission in a nucleus.  Since $\Delta$ is 
not affected at all by the Pauli blocking at hadron level, Fock 
states containing $\Delta$ would not modify the asymmetry in 
a nucleus. However, the nuclear medium effect through the $\Delta$-hole 
excitation would change the pion fields~\cite{dh} although such an 
effect may not be large in light nuclei (like $A = 3$ or $7$ system). 

In this short note we have considered 
the nuclear effect on the flavour asymmetry of anti-quark distributions in a 
bound nucleon in terms of the meson cloud 
model and the Pauli exclusion principle. It is expected that the flavour
asymmetry is enhanced in a nucleus.

\end{document}